# Homogeneous bilayer graphene film based flexible transparent conductor

**Seunghyun Lee,**[a] **Kyunghoon Lee,**[a] **Chang-Hua Liu**[a] **and Zhaohui Zhong** *[a]

[a] Department of Electrical Engineering and Computer Science,
University of Michigan, 1301 Beal Avenue Ann Arbor, MI 48109-2122 U.S.A.; Tel: +01-734-647-1953; E-mail: zzhong@umich.edu

Graphene is considered a promising candidate to replace conventional transparent conductors due to its low opacity, high carrier mobility and flexible structure. Multi-layer graphene or stacked single layer graphenes have been investigated in the past but both have their drawbacks. The uniformity of multi-layer graphene is still questionable, and single layer graphene stacks require many transfer processes to achieve sufficiently low sheet resistance. In this work, bilayer graphene film grown with low pressure chemical vapor deposition was used as a transparent conductor for the first time. The technique was demonstrated to be highly efficient in fabricating a conductive and uniform transparent conductor compared to multi-layer or single layer graphene. Four transfers of bilayer graphene yielded a transparent conducting film with a sheet resistance of 180 $\Omega_\square$ at a transmittance of 83%. In addition, bilayer graphene films transferred onto plastic substrate showed remarkable robustness against bending, with sheet resistance change less than 15% at 2.14% strain, a 20-fold improvement over commercial indium oxide films.

## Introduction

Single and few-layer graphene have emerged as promising materials for novel applications in electronics due to their remarkable optical and electrical properties.[1-5] Their semi-metallic nature with high carrier mobility and low opacity also make them ideal candidates as transparent conductors (TC) for photovoltaic devices, touch panels, and displays.[4,6-8] Indium tin oxide (ITO) is commonly used as a transparent conductor for these applications, but ITO suffers from high cost, material deterioration from ion diffusion, and brittleness making it incompatible with flexible substrates.[7,9] Graphene, on the other hand, shows great promise as a transparent conductor due to its high chemical resistivity, low manufacturing cost, and atomically thin, flexible structure.[2,5,7,9,10]

Several methods have been pursued to synthesize graphene film including reduction of graphene oxide[11-15], liquid exfoliation using organic solvents[16,17], and chemical vapor deposition(CVD).[9,18-25] The CVD method in particular, has drawn great attention as this method yields high quality graphene film. Homogeneous single layer graphene (SLG) can be synthesized on a transition metal substrates with low carbon solubility (e.g. copper) using low pressure CVD (LPCVD).[8,21,22] However, the sheet resistance of a pristine (undoped) SLG is still too large (2000-6000 $\Omega$)[8,9,16] for it to be used as a transparent conductor. Hence, several groups have reported the SLG stacking method with layer-by-layer doping to achieve lower sheet resistance.[8,9,22] The drawback with this approach is that it requires a multitude of transfer processes, which increases the processing time and cost. Alternatively, multi-layer graphene (MLG) with lower sheet resistance can be directly synthesized using LPCVD method on transition metals with relatively high carbon solubility (e.g. nickel),[18-20,26-28] or on copper substrate using atmospheric pressure CVD (APCVD) method.[23,29] However it suffers from several drawbacks such as poor thickness uniformity[18,20,23,26-29] compared to LPCVD grown SLG. Fluctuation of graphene thickness will cause the sheet resistance and the transmittance to vary among different areas of the sample. There was also a report on higher level of defect on APCVD grown MLG compared to LPCVD SLG because of particulate deposition resulting from atmospheric process condition.[23] Furthermore, the MLG method eliminates the possibility of layer-by-layer doping used in a stacked SLG layer, which has been proven to lower the total sheet resistance dramatically.[22]

To this end, we report the use of homogeneous bilayer graphene (BLG) films for a flexible transparent conductor for the first time. The BLG films are synthesized using LPCVD on a copper substrate.[25] In contrast to CVD grown MLG, the BLG film shows high uniformity and very low defect level.[25] By producing uniform, defect free stacks, we demonstrate a BLG based transparent conductor with 180$\Omega_\square$ sheet resistance at 83% transmittance. The use of homogeneous BLG films drastically reduces the processing cost and time compared to SLG based transparent conductors while maintaining high uniformity and quality.

## Experimental section

**Preparation of bilayer graphene based transparent conductor.**

25µm thick copper foil (99.8%, Alfa Aesar) was loaded into an inner quartz tube inside a 3 inch horizontal tube furnace of a commercial CVD system (First Nano EasyTube 3000). The system was purged with argon gas and evacuated to a vacuum of 0.1 Torr. The sample was then heated to 1000°C in H$_2$ (100 sccm) environment with vacuum level of 0.35 Torr. When 1000°C is reached, 70 sccm of CH4 is flowed for 15 minutes at vacuum level of 0.45 Torr. The sample is then cooled slowly to room temperature. The vacuum level is maintained at 0.5 Torr with 100 sccm of argon gas flowing during cooling.

After the CVD synthesis, one side of the copper sample with bilayer graphene is coated with 950PMMA A2 (Microchem) resist and cured at 180°C for 1 minutes. The other side of the sample is exposed to $O_2$ plasma for 30 seconds to remove the graphene on that side. The sample is then left in iron (III) nitrate (Sigma Aldrich) solution (0.05g/ml) for at least 12 hours to completely dissolve away the copper layer. The sample is transferred on to a glass or PET substrate. The PMMA coating is removed with acetone and the substrate is rinsed with deionized water several times. In order to p-dope the sample, graphene on the substrate was immersed in 47.6% nitric acid for 12 hours. The transfer process was repeated several times to create multiple layers of graphene.

**Raman spectroscopy**

The graphene samples were transferred to silicon substrate with 300nm thick $SiO_2$. Raman spectra were collected using a Renishaw inVia Raman Microscopy system equipped with a 17 mW 633 nm He-Ne laser, an 1800 lines/mm grating, and a 20× SLMPlan objective (0.35 numerical aperture). During collection, the slit width was kept at 50 µm and the scanning range was between 1300 and 2900 $cm^{-1}$.

**Transmittance measurement**

Transmittance measurement setup consists of a monochromator (Acton SP2300 triple grating monochromator/spectrograph, Princeton Instruments) coupled with a 250 W tungsten halogen lamp (Hamatsu), a collimator, and a photodetector. Optical filter was used to eliminate higher order diffraction from monochromator. Iris was used to prevent photodector from absorbing the scattered light from the glass substrate. Optical power measurements were carried out using a 1928-C power meter (Newport) coupled to a UV enhanced 918UV Si photodetector (Newport). A blank glass substrate was used as a reference for substraction.

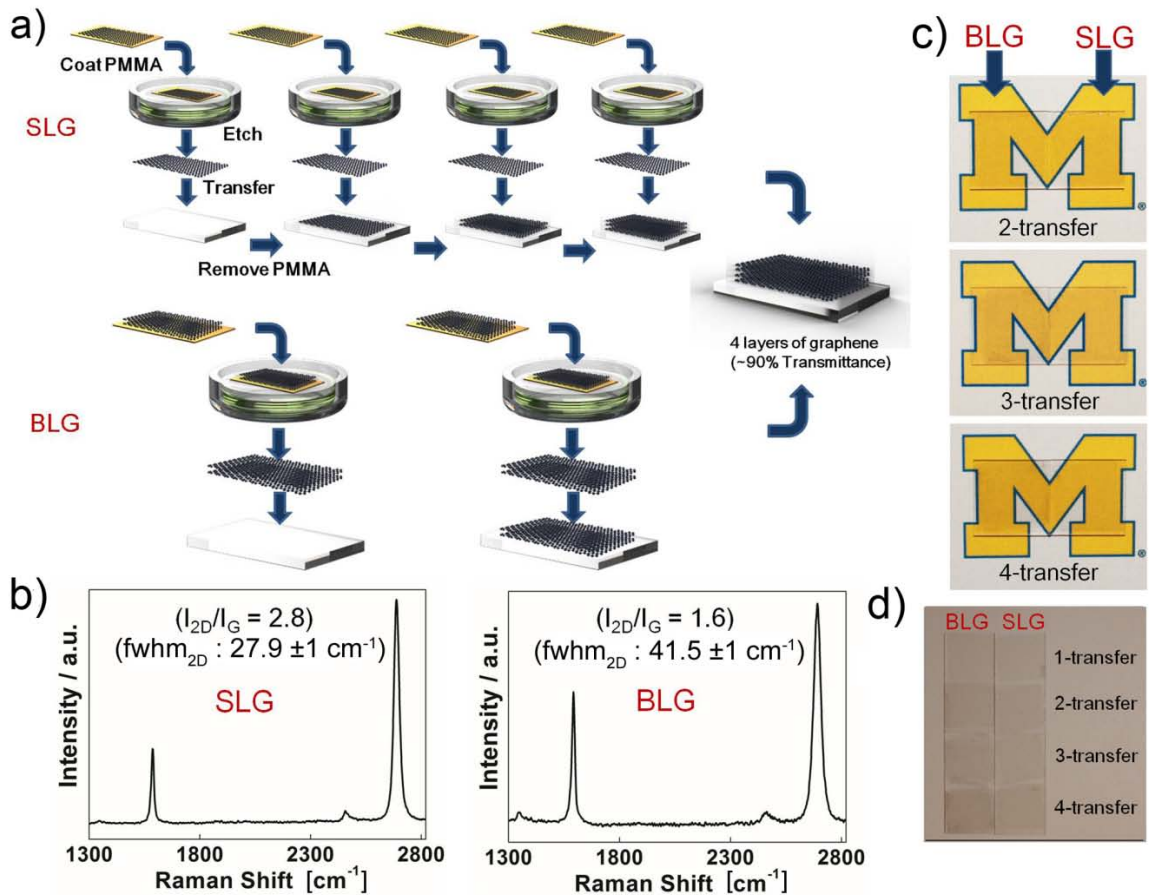

**Fig. 1** (a) Schematic comparison of SLG method and BLG method to synthesize 4 layers of graphene stack to achieve lower sheet resistance. (b) Raman spectra taken from CVD grown SLG (left) and BLG (right) samples. The average values of I2D/IG and fwhm2D from 10 random areas are shown in the plot. (c),(d) Optical comparison of SLG and BLG graphene stacks on glass substrate for 1,2,3,4 transfers with (c) and without(d) background color.

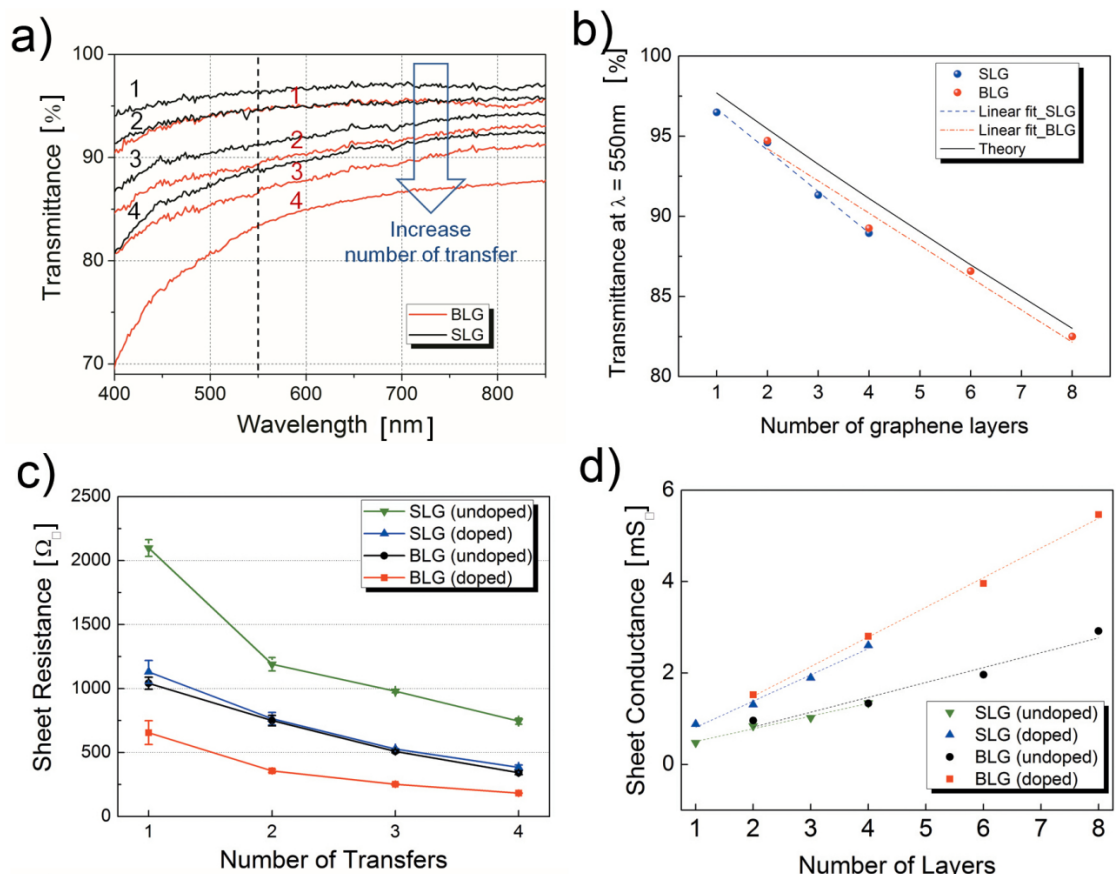

**Fig. 2** (a) Transmittance curve as a function of wavelength for both SLG and BLG stack after 1,2,3,4 transfers respectively. The number near each measurement line indicates the number of transfers. (b) Transmittance value at λ=550nm as a function of graphene layers for SLG and BLG stacks and its linear fits. (c) Sheet resistance of both undoped and doped SLG,BLG stacks with different number of transfers. (d) Sheet conductance of
5 both undoped and doped SLG,BLG stacks as a function of graphene layer number.

**Sheet resistance measurement**

Miller FPP-5000 4-Point Probe Resistivity Meter was used to measure sheet resistance of graphene stacks. For graphene stacks under bending conditions, graphene stacks were first
10 transferred onto PET substrate. Indium oxide sample on 200μm thick PET was purchased(Delta Technology Limited, PF-65IN-1502). Subsequent metal patterning and graphene patterning were done to allow four probe Van der Paaw method while the substrate was bended. The current and
15 voltage difference was measured using a DAQ (National Instruments) in series with a current pre-amplifier (Ithaco, DL instruments 1211).

**Results and discussions**

**Comparison of SLG and BLG stacks**

20 Fig. 1a is an illustration showing a stack of four uniform graphene layers prepared by two different methods using either SLG or BLG. Each transfer process consists of multiple steps that include CVD synthesis, coating of graphene with poly methyl methacrylate (PMMA), copper etching, transfer,
25 drying, and removing of the polymer layer. In order to form a stack of four graphene layers, four repeated transfers are needed when using SLG, while only two transfers are required for BLG. It is clear that the BLG method significantly reduces the amount of raw materials and time
30 required by reducing the number of transfer processes by half.

Raman spectra were taken at 10 random spots on the CVD graphene films to verify the number of graphene layers for both SLG and BLG (Fig. 1b). The two most important parameters in determining SLG and BLG from the Raman
35 spectra are the ratio of 2D band (~2691 cm$^{-1}$) intensity to G band (~1595 cm$^{-1}$) intensity ($I_{2D}/I_G$) and the full width at half maximum (fwhm$_{2D}$) value of the 2D band.[30,31] The mean value of the $I_{2D}/I_G$ ratio is 2.8 for SLG and 1.6 for BLG, while the mean value of the 2D band fwhm$_{2D}$ is 27.9 cm$^{-1}$ for SLG
40 and 41.5 cm$^{-1}$ for BLG. These Raman spectra values are definitive indications of SLG and BLG, respectively. The differences in the opacity of SLG versus BLG stacks become more obvious as the number of transfers increases (Fig. 1c). This is because the difference in the number of graphene
45 layers increases from two to four layers as the number of transfers increases from two to four. Fig. 1d shows the direct optical comparison of both SLG stacks and BLG stacks without the background color.

## Transmittance and sheet resistance

Furthermore, we measured the transmittance (T) of both SLG stacks and BLG stacks on glass substrates for comparison (Fig. 2a). It is clear that the transmittance of both SLG and BLG stacks drops as the number of transfers increases. For quantitative comparison, the transmittance values of SLG 1-, 2-, 3-, and 4-transfer stacks at 550nm wavelength[8,9,19,22] are measured to be 96.5%, 94.6%, 91.3%, and 89.0%, respectively. The transmittance values of BLG 1-, 2-, 3-, and 4-transfer stacks at 550nm wavelength are 94.7%, 89.3%, 86.6%, and 83.0%, respectively. This result indicates that as expected, BLG's opacity is twice the value of SLG. The transmittance spectrum decreases as it nears the ultraviolet region due to exciton-shifted Van Hove singularity in the graphene density of states.[5] It is also interesting to note that the downward shift in transmittance near the high energy region is more significant as the number of stacked layers increases. This was observed in many other works[8,9,20,22] and it may be due to residue trapped between layers.

Fig. 2b shows the transmittance values at 550 nm as a function of the total graphene layer numbers, and compares them with the theory. Nair et al have shown that transmittance of graphene is defined by the fine structure constant $\alpha \approx 0.0073$ and the transmittance of single graphene layer can be expressed as $T \approx 1-\pi\alpha \approx 97.7 \pm 0.1\%$.[6] Hence, the transmittance of multiple layers can be expressed as $T^n = (1-\pi\alpha)^n$, where n is the number of layers.[32] The plots confirm that the increases in opacity of both BLG stacks and SLG stacks are close to the theoretical value. The offset of 1%-2% from the theory can be observed and we believe the deviation is likely due to a small amount of polymer residue (e.g. PMMA) that may have been trapped between the sandwiched layers.

We also characterized the sheet resistance ($R_\square$) values for both undoped and nitric acid doped SLG and BLG stacks using four probe method (Fig. 2c). Each data point is taken from 10 different regions on each sample and standard deviation values are expressed with error bars. As the number of transfers increases, the sheet resistance decreases for both doped and undoped samples. The sheet resistance also drops roughly by a factor of two after layer-by-layer nitric acid doping.

The total resistance of multiple layers of graphene is composed of both in-plane sheet resistance of individual layers and inter-layer resistance between layers.[9] High inter-layer resistance implies resistive interface that will cause most of the current to flow only at the top most layer.[9] To investigate the effect of inter-layer resistance in multi-layer graphene stacks, we plot sheet conductance $G_\square$ versus the number of graphene layers in Fig. 2d. Linear fits for undoped samples yield 0.278 mS$_\square$/layer for SLG stacks and 0.325 mS$_\square$/layer for BLG stacks, which shows a 17% increase for BLG stacks. Linear fit for doped samples yielded a 0.574 mS$_\square$/layer for SLG stacks, and 0.649 mS$_\square$/layer for BLG stacks which shows a 13% increase for BLG stacks. It is interesting to note that sheet conductance per layer for BLG was found to be slightly higher than that of SLG. The result is unexpected because SLG based conductors have been doped twice as many times compared BLG based conductors. It is known that a randomly stacked graphene structure will have large interlayer distances that would strongly reduce the electronic dispersion perpendicular to the basal plane compared to a Bernal-like or an ordered stack structure.[33,34] Since a SLG stack consists of only randomly stacked layers while a BLG stack will retain its ordered layers between each transferred layer, it is possible that BLG was advantageous in maintaining stronger coupling between adjacent layers. In addition, the number of interfaces created from the transfer processes is lower for BLG compared to that of SLG. For example, a four graphene layer stack consists of three transfer interfaces for a SLG stack while only one transfer interface exists in a BLG stack. This may have also helped in lowering the total inter-layer resistance.

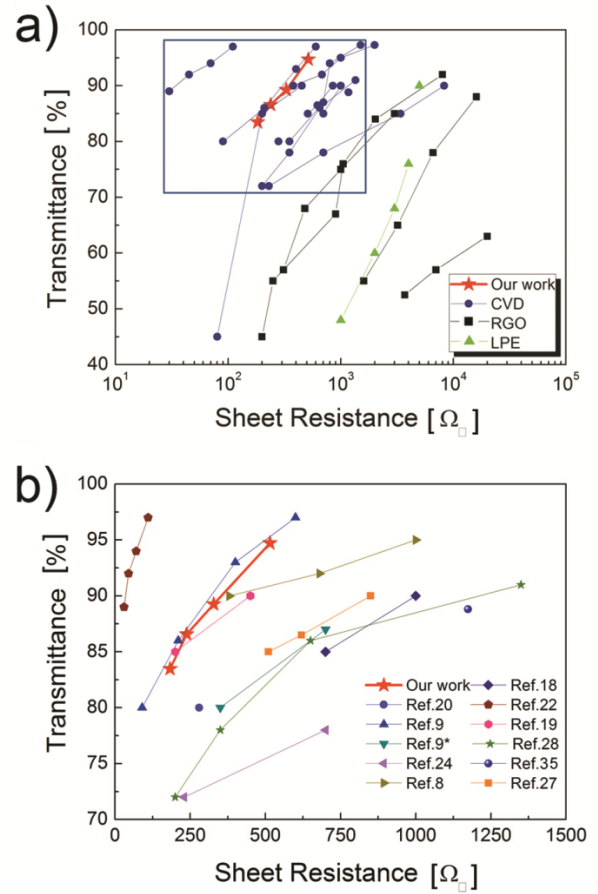

**Fig. 3** (a),(b) Transmittance versus sheet resistance for graphene based transparent conductors grouped according to production methods in log scale (a) and only with CVD method in linear scale(b). Blue rectangle in (a) represents the range of x,y axis for (b). Ref 9* in (b) is the value for undoped graphene in reference 9.

## Comparison with other methods

Recent progress in graphene transparent conductor, in terms of transmittance and sheet resistance, is summarized in Fig. 3a and 3b. In Fig. 3a, the reports are categorized according to different production strategies. The quality of transparent conductor is superior as the characteristics line leans toward the upper left region of the graph, indicating a higher transmittance with lower sheet resistance.[5,7] In most cases, CVD grown graphene[9,18-22,24,26-28,35] has been proven to be superior compared to liquid based synthesis method such as reduction of graphene oxide (RGO)[11-14] and liquid phase exfoliation (LPE)[16,17] due to its inherent lack of structural defect.[5,7] Fig. 3b focuses only on CVD methods with nitric acid as the dopant and the sheet resistance is shown with the x-axis as the linear increment. Our results using BLG are comparable or better than other CVD methods using SLG stacks or MLG.

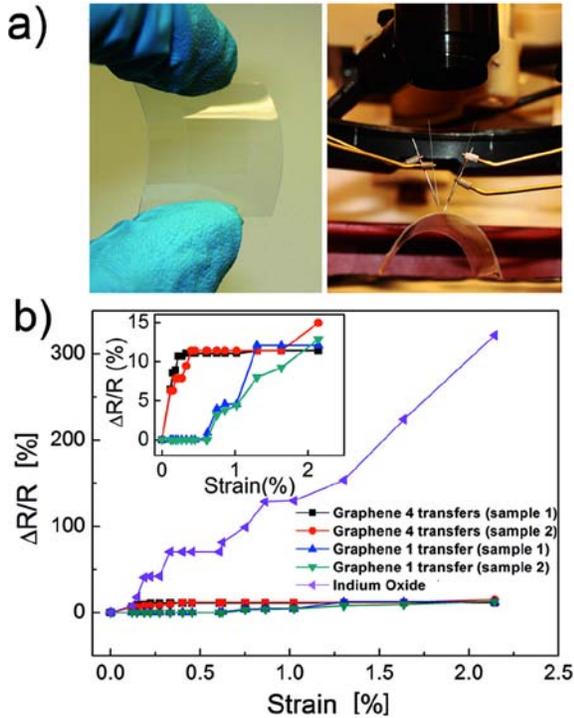

**Fig. 4** a) Photographs of graphene film on flexed PET substrate(left) and measurement setup of strained substrates (right). (b) Variation in resistance of stacked BLG films and indium oxide films on 200μm thick PET substrate as a function of strain values.

**Sheet resistance change with strain**

One of the most significant advantages of a graphene based transparent conductor is its electromechanical stability and mechanical flexibility.[7,8,13,20,22,24] To test the sheet resistance of BLG stacks under a mechanical deformation, we transferred BLG films onto 200μm thick polyethylene terephthalate (PET) flexible substrates and patterned them with gold electrodes for four-probe measurement (Fig. 4a). Two samples of both BLG 1-transfer and BLG 4-transfer were tested in comparison with a commercial Indium oxide on a PET substrate under bending condition. Fig. 4b shows relative change in sheet resistance versus strain due to bending. The radius of curvature is converted to the unit of strain from the equation $\varepsilon = d/2r$, where $\varepsilon$ is surface strain, $d$ is substrate thickness, and $r$ is radius of curvature.[36] At 2.14% strain, the sheet resistance of the indium oxide sample increased by 321% while the graphene samples only increased by 10 to 15%. The indium oxide sample shows a drastic change in sheet resistance due to its brittle nature while graphene samples are much more robust against bending. The inset of Fig. 4b shows a more detailed comparison between BLG 1-transfer and BLG 4-transfer samples. It is interesting to note that BLG 4-transfer samples show slight increase in sheet resistance (~10%) at a lower strain than BLG 1-transfer samples. This was not reported in any previous literature. The shear stress that acts between the stacked layers[37] may have disrupted the interface state between graphene layers, and bending the substrate may increase the inter-layer resistance leading to earlier increase in the sheet resistance. Detailed understanding will require further study.

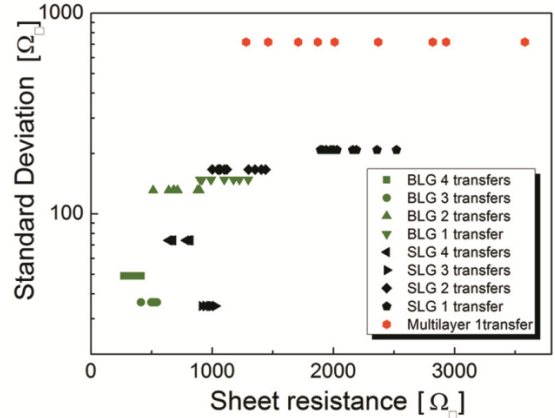

**Fig. 5** Distribution of sheet resistance and its standard deviation values for SLG, BLG stacks and a cvd grown multilayer (MLG) sample. 10 measurements were taken on different areas of each sample.

**Uniformity of BLG stacks**

Last, we evaluate the uniformity of graphene film stacks by taking ten measurement points from different areas of each sample. In Fig. 5, the sheet resistance values of SLG, BLG stacks, and MLG are plotted with standard deviation as the vertical axis. The actual values of sheet resistance are plotted for facile observation of the distribution. The MLG sample was grown with the APCVD method (Fig.S1†) on copper substrate. BLG and SLG samples show similar distribution of sheet resistance and standard deviation for films with the same number of transfers. On the other hand, the MLG sample shows high standard deviation indicating a higher level of non-uniformity in sheet resistance across the sample area. The result agrees with several publications reporting non-uniformity in thickness for MLG.[18-20,23,26,29] We also acknowledge that the sheet resistance of MLG has strong correlation with surface roughness[29] and there is an effort to produce a smoother (more uniform) multilayer graphene. Nonetheless, BLG stacks stand out with both better uniformity than MLG and drastically reduced fabrication complexity compared to SLG stacks. It is also interesting to note that the standard deviation value becomes lower as the number of stacks increases. This may be attributed to the increased number of graphene layers that can act as channels to negate certain high resistivity areas (e.g. wrinkles or defects) that may reside on one of the layers in the stack.

## Conclusions

SLG stacks have been proven to be a high quality transparent conductor in many reports.[8,9,22] However, most researches overlook the fact that SLG stacks require multiple graphene transfers that results in considerable amount of material waste due to metal wet etching. Furthermore, transferring a large area of graphene is a delicate process that may jeopardize the overall quality of graphene and it is best to minimize the number of transfers. Our BLG method can significantly simplify the process to save cost, time, and reduce waste. Furthermore, the quality and uniformity of BLG stack based transparent conductors have been confirmed to be very high. Although our method of nitric acid doping lowered the sheet resistance by a factor of two, using different dopants and doping methods can lead to further reduction of sheet resistance by a factor of three to five.[9,32] Utilization of a graphene hybrid structure[38] with BLG can also open up new possibilities for an ultra-low sheet resistance transparent conductor. Lastly, the size of our BLG film is only limited by the synthesis apparatus and can be readily scaled up, thus enabling applications with large area flexible transparent conductors.

## Acknowledgment

Acknowledgment is made to the Donors of the American Chemical Society Petroleum Research Fund, the U-M/SJTU Collaborative Research Program in Renewable Energy Science and Technology, and National Science Foundation Scalable Nanomanufacturing Program (DMR-1120187). This work used the Lurie Nanofabrication Facility at University of Michigan, a member of the National Nanotechnology Infrastructure Network funded by the National Science Foundation.

8Electronic Supplementary Information for

# Homogeneous Bilayer Graphene Film Based Flexible Transparent Conductor

Seunghyun Lee,[a] Kyunghoon Lee,[a] Chang-Hua Liu[a] and Zhaohui Zhong *[a]

[a] Department of Electrical Engineering and Computer Science,
University of Michigan, 1301 Beal Avenue Ann Arbor, MI 48109-2122 U.S.A.; Tel: +01-734-647-1953; E-mail: zzhong@umich.edu**Multilayer graphene (MLG) characterization and growth**

25µm thick copper foil (99.8%, Alfa Aesar) was loaded into an inner quartz tube inside a 3 inch horizontal tube furnace of a commercial CVD system (First Nano EasyTube 3000). The system was purged with argon gas and evacuated to a vacuum of 0.1 Torr. The sample was then heated to 1000°C with argon (1000sccm) and hydrogen (50 sccm) flow at atmospheric pressure for annealing. When 1000°C is reached, the annealing process is maintained for 30 minute, and then 50 sccm of $CH_4$ is flowed for 5 minutes at atmospheric pressure. The sample is then cooled to room temperature without $CH_4$ gas flow. The hydrogen is cut off but the argon flow is maintained during the cooling process. The time plot of the entire growth process is shown in Fig. S1a. After the transfer process to silicon substrate with thermal oxide, Raman spectroscopy was used to verify the existence of multilayer graphene. For this typical raman spectra shown in figure S1b, $I_{2D}/I_G$ ratio is 0.78 and the $fwhm_{2D}$ is 63. However for 10 different measurements on random areas, the value of $I_{2D}/I_G$ varied from 0.51 to 1.11 and $fwhm_{2D}$ varied from 47.78 to 68.27.

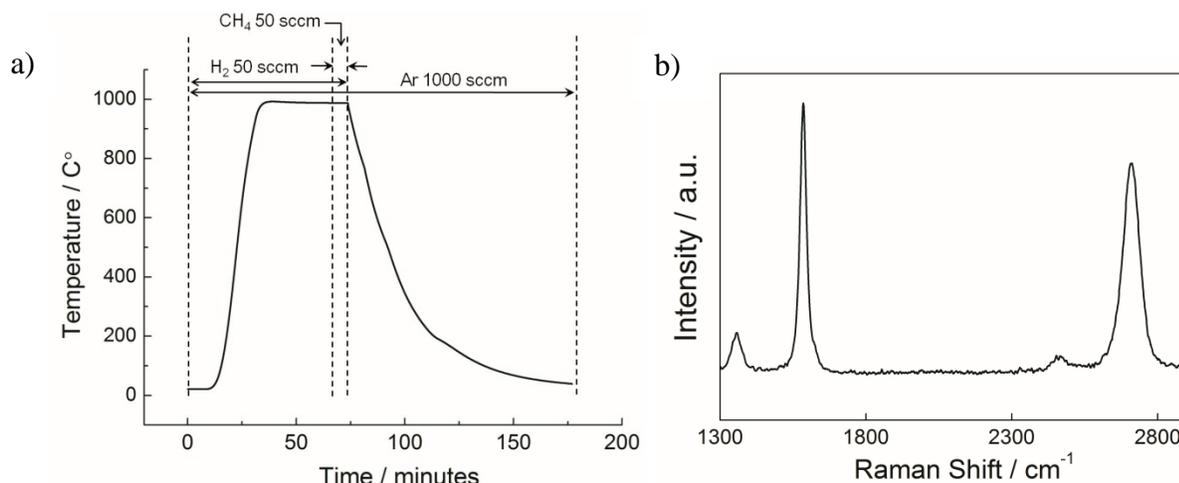

**Figure S1.** a) Temperature vs. time plot of multilayer graphene growth condition. Pressure is maintained to atmospheric pressure at all time except the initial purge stage. b) Raman spectroscopy result showing typical multilayer signal.

S1